\documentclass[11pt]{article}
\usepackage{graphicx}
\usepackage{booktabs}
\usepackage{multirow}
\usepackage{mathalfa} 
\usepackage{authblk}
\usepackage[flushleft]{threeparttable}
\usepackage{natbib}
\usepackage{centernot}
\usepackage{xcolor}
\usepackage{amssymb}
\usepackage{pifont}
\usepackage{bm}
\usepackage{mathalfa} 
\usepackage{amsmath}

\renewcommand{\baselinestretch}{1.5}

\newcommand {\Var} {\mbox{Var} }   
\newcommand {\Power} {\mbox{Power} }

\def\bSig\mathbf{\Sigma}

\usepackage{chngcntr}
\counterwithout{figure}{section}

\begin{document}

\title{\bf
Testing for Treatment Effect Twice Using Internal and External Controls in Clinical Trials
}

\author{Yanyao Yi$^*$, Ying Zhang$^*$, Yu Du}
\affil{Global Statistical Sciences, Eli Lilly and Company, Indianapolis, Indiana 46285, U.S.A.\thanks{equal contribution}}

\author{Ting Ye}
\affil{Department of Biostatistics, University of Washington,  Seattle, Washington, 98195, U.S.A.\thanks{tingye1@uw.edu}}

\maketitle

\begin{abstract}
  {Leveraging external controls -- relevant individual patient data under control from external trials or real-world data -- has the potential to reduce the cost of randomized controlled trials (RCTs) while increasing  the proportion of trial patients given access to novel treatments. However, due to lack of randomization, RCT patients and external controls may differ with respect to covariates that may or may not have been measured. Hence, after controlling for measured covariates, for instance by matching, testing for treatment effect using external controls may still be subject to unmeasured biases.  In this paper, we propose a sensitivity analysis approach to quantify the magnitude of unmeasured bias that would {be needed} to alter the study conclusion that presumed no unmeasured biases are introduced by employing external controls. Whether leveraging external controls increases power or not depends on the interplay between sample sizes and the magnitude of treatment effect and unmeasured biases, which may be difficult to anticipate. This motivates a combined testing procedure that performs two highly correlated analyses, one with and one without external controls, {with a small correction for multiple testing} using the joint distribution of the two test statistics. {\color{black} The combined test provides a new method of sensitivity analysis designed for data fusion problems, which anchors at the unbiased analysis based on RCT only and spends a small proportion of the type I error to also test using the external controls. {In this way, if leveraging external controls increases power, the power gain compared to the analysis based on RCT only can be substantial; if not, the power loss is small.}
		The proposed method is evaluated in theory and power calculations, and applied to a real trial.}}
\end{abstract}


\section{Introduction}
\label{sec: intro}

\subsection{Use of external controls in randomized controlled trials}

Randomized controlled trials (RCTs) are the gold standard for generating high-quality causal evidence of new treatments and have long been recognized as the standard method to support key decisions in the drug development process \citep{jones2015history, bothwell2016emergence}. However, despite its clear advantages, the traditional paradiam of conducting RCTs has been increasingly criticized for failing to meet contemporary needs. In certain settings, for example, in HIV prevention  \citep{JANES2019e475, sugarman2021ethical}, oncology \citep{rahman2021leveraging}, and neurology \citep{mintzer2015separate}, randomizing patients to placebo may be difficult for ethical or feasibility reasons. Moreover, adequately powered RCTs are becoming more and more impractical as a growing number of new treatments are targeted toward rare diseases or biomarker-defined subgroups of patients in the era of precision medicine \citep{eichler2021randomized}. 

Meanwhile, a plethora of real-world data (RWD) have been curated for administrative or research purposes and are becoming accessible to researchers in the form of disease registries, administrative claims databases, and electronic health records. These rich data sources can produce valuable insights, i.e., real-world evidence (RWE), into the effect of treatments in routine, daily practice. However, researchers almost ubiquitously caution against possible bias from unmeasured confounding when using RWD.

Being well aware of the limitations of using either RCT or RWD alone, the idea of using RWD to supplement RCT has gained growing interest in recent years. As forcefully argued in \citet{eichler2021randomized}, ``the future is not about RCTs vs. RWE but RCTs and RWE.''  There are numerous opportunities in how the integration of RCTs and RWD can achieve fruitful results that using either RCT or RWD alone can not \citep{colnet2020causal, degtiar2021review,shi2021data}. Among those, an important theme is on augmenting the RCT with RWD to increase efficiency \citep{yang2020elastic, yang2020improved, gagnon2021precise, chen2021minimax, cheng2021adaptive, li2021efficient}, and particularly, constructing an externally augmented control arm in the analysis of RCTs \citep{li2020improving, harton2021combining, gao2021sample, liu2022matching}. Leveraging external controls -- relevant individual patient data under control from external trials or real-world data --   has the potential to reduce the cost of RCTs while increasing  the proportion of trial patients given access to novel treatments.


Using external controls is not an entirely new idea. Criteria for evaluating what constitute an acceptable external control arm are proposed in \citet{Pocock1976}. 
It was discussed twenty years ago by the \citet[E10 Section 2.5]{iche10}, and also recognized by the \citet{ema2006}, \citet{fda2018}, and National Cancer Institute \citep{Sharpless2019} as one direction to modernize clinical trials.  In fact, properly selected external controls (e.g., using propensity score matching) have shown early promise, and several drugs have already been approved based on external control groups \citep{carrigan2020using, schmidli2020beyond, thorlund2020synthetic}. 

Using external controls typically requires the exchangeability condition, i.e., all patient characteristics that affect the potential outcome under control and differ between the trial population and the external control population are measured \citep{Stuart:2011aa}. While careful adjustment for observed covariates can probably render the exchangeability assumption to hold approximately, the analysis may still be biased due to unmeasured covariates related to ``difficulties in reliably selecting a comparable population because of potential changes in medical practice, lack of standardized diagnostic criteria or equivalent outcome measures, and variability in follow-up procedures'' \citep{fda2018}. {\color{black} 
	To reduce the potential biases from using external controls,
	an intuitive frequentist approach is ``test-then-pool''  that first tests for the comparability of the external controls and internal controls before leveraging external controls \cite{liu2022matching}. Bayesian methods that rely on power priors have also been popular, which use the likelihood of the external data to a specified power as the prior distribution \cite{Chen&Ibrahim2000, nikolakopoulos2018dynamic}. As such, one can use power priors to adjust the weight allocated to the external information according to the levels of comparability between the external control and the internal data. However, these methods lack formal statistical theory on how the unmeasured biases might affect the validity and efficiency of the proposed procedures. 
	
	{In this article, we take a different perspective to this problem and propose a sensitivity analysis approach to quantify the magnitude of unmeasured bias
		that would be needed to alter the study conclusion that presumed no unmeasured biases are introduced by
		employing external controls \citep{iche9r1}. With the unbiased RCT-only test as the benchmark, leveraging external controls increases power or not depends on the interplay between sample sizes and the magnitude of treatment effect and unmeasured biases, which may be difficult to anticipate. This motivates a combined testing procedure that performs both tests, one with and one without external controls, correcting for multiple testing using the joint distribution of the two test statistics. Because the two tests are highly correlated, this correction for multiple testing is small. Interestingly, the proposed combined testing procedure can be viewed as a new method of sensitivity analysis designed for data fusion problems that anchors at the unbiased analysis based on RCT only and ``spends'' a small proportion of the type I error (i.e., the cost of multiple testing) to also test using the pooled controls. In this way, if leveraging external controls increases power, the power gain compared to the RCT-only test can be substantial; if not, the power loss is small.
}} Before introducing technical details, it is useful to consider a motivating example. 

\subsection{Example: a randomized controlled trial in patients with type-2 diabetes}
\label{subsec: motivating example}

Consider a non-inferiority, phase 3  RCT (referred to as the internal trial,  ClinicalTrials.gov number, NCT01894568) comparing a new basal insulin, insulin peglispro,
to insulin glargine as the control in Asian insulin-na{\"\i}ve patients with type-2 diabetes using a noninferiority margin of 0.4\% \citep{hirose2018open}. The primary endpoint is the change in hemoglobin A1c (HbA1c) from baseline to 26 weeks of treatment. HbA1c is a continuous-valued measure of average blood glucose in the past three months.
Before this trial, a phase 3 RCT of similar design (referred to as the external trial, ClinicalTrials.gov number, NCT01435616) has been conducted in the North America and Europe \citep{davies2016basal}, whose control arm will be used as the source of external controls.

We focus on the overweight and obese population, which are respectively defined as $23\leq$ Body Mass Index (BMI) $<25$ and $ \text{BMI} \geq 25 $ for the internal trial according to the Asia-Pacific guidelines, and $25\leq \text{BMI}<30$  and $\text{BMI} \geq 30 $ for the external trial according to the World Health Organization classifications \citep{lim2017comparison}. There are in total 159 patients under treatment and 150 patients under control in the internal RCT, and 486 patients under control in the external trial.
We match 159 similar external controls to the 159 treated patients in the internal RCT using optimal matching based on a robust Mahalanobis distance and a caliper on the propensity score. See \citet[Part II]{rosenbaum2020design} for discussion of these matching techniques. Table \ref{tb: table1} describes covariate balance in the 159 matched pairs. All variables have standardized differences less than $0.13$ and are considered sufficiently balanced \citep{rosenbaum2002observational}.

Using only the internal RCT, 159 patients under treatment and 150 under control, we conduct a $Z$-test with the noninferiority margin of 0.4\% and obtain a one-sided p-value $7.92\times 10^{-7}$. {In this analysis, the evidence that the new insulin treatment is noninferior to insulin glargine is strong enough when only using the internal controls.} On the other hand, under the exchangeability assumption, which implies that the 159 matched external controls are comparable to patients in the internal RCT, we construct an augmented control arm of 309 patients in total and obtain a one-sided p-value $1.88\times 10^{-7}$. Again, we find strong evidence of noninferiority; however, an investigator may be in doubt about the exchangeability assumption due to the influence of regions on the outcome. Then a natural question is could the one-sided p-value of $1.88\times 10^{-7}$ be due to regions rather than the effect of treatment? {\color{black} If the study conclusion from using external controls can be altered by a plausible effect of regions  and because the RCT-only test is already powerful enough, the RCT-only test would be a better choice. However, it would be difficult to know this before examining the data. Motivated by the advice of performing multiple analyses with an appropriate correction for multiple testing given by \citet{rosenbaum2012testing}, we propose a combined testing procedure that performs both analyses, controlling for multiple testing using the joint distribution of the two test statistics. In this article, we will demonstrate that the combined test avoids making an inapt choice about whether to use external controls or not, and only has a small loss of power compared to knowing a priori which is the better choice.
}


\subsection{Outline}
Section \ref{sec: testing} presents a test that uses only the internal controls and another test that also leverages the external controls, and discusses controlling type I error and comparing power without the exchangeability assumption. Section \ref{sec: combined test} proposes a combined test that performs both tests and studies in detail its statistical properties. Section \ref{sec: simu} presents power calculations. Section \ref{sec: real} returns to the real data applications. Section \ref{sec: discuss} 
concludes with a discussion.

\section{Testing Using Internal and External Controls}
\label{sec: testing}
\subsection{Testing Under Exchangeability}
\label{subsec: testing Exchangeability}

There is a randomized controlled trial (RCT) denoted as $D=1$. Let $A$ be a binary treatment, where $A=1$ denotes treatment and $A=0$ denotes control, $X$ a vector of  observed baseline covariates, $Y^{(a)}$ the potential outcome under $A=a$, for $a=0,1$. Throughout the article, we assume consistency and Stable Unit Treatment Value Assumption (SUTVA) so that the observed outcome satisfies $Y= AY^{(1)} + (1-A) Y^{(0)}$  \citep{Rubin1980}. Our estimand of interest is the average treatment effect in the RCT population $\theta^\star = E(Y^{(1)}\mid D=1) - E(Y^{(0)}\mid D=1)  $. In particular, we consider testing a one-sided hypothesis: 
$$H_0: \theta^\star = \theta_0 \quad \text{versus}\quad H_A: \theta^\star > \theta_0.$$
The other direction can be considered in the same way. Combining both one-sided tests and applying Bonferroni correction give a two-sided test \citep[Section 4.2]{cox1977role}, and by inversion, a confidence interval.

Write the RCT sample as $(Y_i,  X_i, A_i, D_i=1), i=1,\dots, n_{r}$, which is assumed to be independent and identically distributed according to the joint law of $(Y^{(1)}, Y^{(0)},  X, A)\mid D=1$. Randomization in the RCT guarantees that $A\perp (Y^{(1)}, Y^{(0)}, X) \mid D=1$ and $P(A=a\mid D=1)= \pi_a>0 $ for $a=0,1$, with $\pi_a$ known and $\pi_0+\pi_1= 1$.  Let $\overline{Y}_a$ and  $S_a^2 $  respectively be the sample mean and sample variance of the responses $Y_i$'s from RCT subjects under treatment $a$,  for $a=0,1$. Hence, the null hypothesis $H_0$ can be tested using a simple $Z$-statistic: 
\[
T_1 = \frac{  \overline{Y}_1 - \overline{Y}_0 - \theta_0 }{\sqrt{ n_1^{-1}S_1^2+n_0^{-1}S_0^2}},
\]
where $n_1 $ and $ n_0$ are respectively the number of RCT patients under treatment and control. Based on $T_1$, we reject $H_0$ when $T_1 \geq z_{1-\alpha}$, where $z_{1-\alpha}$ is the $(1-\alpha)$th quantile of the standard normal distribution.

To supplement the RCT using external controls, one approach is to first extract external data for patients under control based on the inclusion/exclusion criteria of the RCT and then proceed by matching these external patients to the RCT patients based on their similarity in observed baseline information $X$ \citep{schmidli2020beyond}. Let $D=0$ denote the matched external controls, {\color{black} and thus $D=0$ implies $A=0$}. Write the matched external controls as $(Y_i,  X_i, A_i=0, D_i=0), i=1,\dots, n_e$, which is assumed to be independent and identically distributed according to the joint law of $(Y^{(0)}, X)\mid A=0, D=0$. Suppose that matching has rendered the baseline observed covariates comparable between the RCT and external controls, i.e., $D\perp X$, and that these baseline covariates $X$ explain all differences between the RCT and external controls, i.e., the exchangeability assumption $D\perp Y^{(0)} \mid X$ holds. This implies $D\perp (Y^{(0)}, X)$ and thus $E(Y^{(0)} \mid D=1 ) = E(Y^{(0)} \mid D=0) $. Let $\overline{Y}_e$ be the sample mean of the responses $Y_i$'s from the external controls, and $ w\overline{Y}_0+ (1-w)\overline{Y}_e$ be a weighted average of mean responses for the two control groups, where $w\in [0,1]$ is a pre-specified weight, which could reflect the proportion of the internal control in the two control groups combined. Therefore, the null hypothesis $H_0$ can also be tested borrowing information from the external controls using 
$$
T_2 (w)= \frac{\overline{Y}_1 -  \{w\overline{Y}_0+ (1-w)\overline{Y}_e \}- \theta_0}{\sqrt{ n_1^{-1}S_1^2 +w^2 n_0^{-1}S_0^2 +(1-w)^2 n_e^{-1} S_e^2 }},
$$
where $S_e^2$ is the sample variance of the responses $Y_i$'s from external controls. We make two remarks about $T_2(w)$. First, $T_2(w)$ is constructed assuming independence between the RCT and external controls, which means that $T_2(w)$ may be  conservative due to correlation induced by matching  \citep{Austin2014} but usually to a small extent as the correlation is typically small \citep{Schafer2008}. Second, $T_2(w), w\in[0,1]$ defines a family of statistics that includes $T_2(1)= T_1$ as a special case. Among those, the exchangeability assumption implies the optimal $w$ that maximizes the efficiency of $T_2 (w)$ is proportional to the sample size, i.e., the optimal $w$ equals $ (n_r\pi_0)/(n_r\pi_0+n_e )$. One can also choose different values of $w$ to reflect the weights allocated to the two control groups.

\subsection{Controlling Type I Error Without Exchangeability}

The aforementioned approach of leveraging external controls relies on the exchangeability assumption, which may not hold because the RCT patients and external controls may differ with respect to covariates that may not have been measured. Without exchangeability, $\overline{Y}_1 - \{w\overline{Y}_0+ (1-w)\overline{Y}_e \}$ is not necessarily centered at $\theta_0$ under $H_0$ and rejecting the null hypothesis when $T_2 (w) \geq z_{1-\alpha}$ may inflate type I error. 

{\color{black}Define $\Delta^\star= E(Y^{(0)}\mid D=1) -  E(Y^{(0)}\mid D=0)$, which may be nonzero when exchangeability does not hold.} {This could occur, for example, if an important prognostic variable is unobserved and left uncontrolled, or if a variable that differs in distribution between $D=0$ and $D=1$ (such as region) cannot be matched.}   The correct rejection region for a size-$\alpha$ test based on  $T_{2} (w)$ is 
\begin{align*}
	T_{2} (w) - \frac{(1-w)\Delta^\star}{\sqrt{ n_1^{-1}S_1^2 +w^2 n_0^{-1}S_0^2 +(1-w)^2 n_e^{-1} S_e^2 }}>z_{1-\alpha},
\end{align*}
which is infeasible because $\Delta^\star$ is unknown. To deal with this issue, a tempting choice is to estimate $\Delta^\star$ by $\overline Y_0 -  \overline{Y}_e$ and adjust the numerator of $T_2(w)$ to make it mean zero. Nonetheless, this ``de-biasing'' step introduces additional variation and the resulting test statistic becomes equivalent to $T_1$, the test statistic without using any external controls. 


In order to borrow information from external controls while still controlling type I error, we consider departures from the exchangeability through the lens of a sensitivity analysis \citep{rosenbaum2020design}. {\color{black}
	Specifically, we consider a sensitivity parameter $\Delta_0$ such that it bounds the magnitude of bias $\Delta^\star$, i.e., 
	$\Delta_0\geq \Delta^\star$. } Define 
$$
T_{2, \Delta_0} (w) = \frac{\overline{Y}_1 - \{w\overline{Y}_0+ (1-w)\overline{Y}_e\} - \theta_0- (1-w){\Delta_0}}{\sqrt{ (n_r \pi_1)^{-1}S_1^2 +w^2 (n_r \pi_0)^{-1}S_0^2 +(1-w)^2 n_e^{-1} S_e^2 }}.
$$
Because $\Delta^\star\leq \Delta_0$,  the reject region $T_{2, \Delta_0} (w)\geq z_{1-\alpha}$ controls type I error at level $\alpha$. As a special case when $\Delta^\star\leq \Delta_0$ holds with $\Delta_0 =0$ (e.g., under exchangeability), $T_{2, \Delta_0} (w)\geq z_{1-\alpha}$ becomes $T_2(w)\geq z_{1-\alpha}$, the reject region under exchangeability. As $\Delta_0$ increases, there is greater uncertainty about how the exchangeability might be violated, leading to more stringent rejection criterion to control type I error. The reject region $T_{2, \Delta_0} (w)\geq z_{1-\alpha}$ is sharp under $\Delta^\star\leq \Delta_0$ in the sense that they are of size-$\alpha$ when $\Delta^\star=\Delta_0$, so it cannot be improved unless further information is provided.

\subsection{Power Comparison  Without Exchangeability}
\label{subsec: power comparison without exchangeability}

Write $\sigma_a^2= \Var(Y^{(a)}\mid D=1)$, for $a=0,1$, and $\sigma_e^2 = \Var(Y^{(0)}\mid D=0)$. Under the alternative hypothesis $H_A: \theta^\star>\theta_0$, the power of $T_1$ is the probability of event $T_1\geq z_{1-\alpha}$, which is asymptotically equal to 
\begin{align}
	1-\Phi\left(z_{1-\alpha}+ \frac{ \sqrt{n_r} (\theta_0 - \theta^\star )}{  \sqrt{ \pi_1^{-1} \sigma_1^2 + \pi_0^{-1} \sigma_0^2 } }\right),   \label{eq: power1}
\end{align}
where $\Phi (\cdot) $ is the standard normal cumulative distribution.  In parallel, the power of $T_{2, \Delta_0} (w)$ is the probability of event $T_{2, \Delta_0} (w)\geq z_{1-\alpha}$, which is asymptotically equal to  
\begin{align}
	1-\Phi\left(z_{1-\alpha}+
	\frac{\sqrt{n_r} (\theta_0 - \theta^\star )+ \sqrt{n_r} (1-w)(\Delta_0-\Delta^\star) }{\sqrt{ \pi_1^{-1}\sigma_1^2 +w^2  \pi_0^{-1}\sigma_0^2 +(1-w)^2n_r n_e^{-1} \sigma_e^2}}\right).   \label{eq: power2}
\end{align}

Several remarks are in order based on the above power formulas.  First, the power of $T_{2, \Delta_0} (w)$ is larger than that of $T_1$ if and only if    
\begin{align*}
	\frac{ \theta_0 - \theta^\star + (1-w)(\Delta_0-\Delta^\star) }{\sqrt{ \pi_1^{-1}\sigma_1^2 +w^2  \pi_0^{-1}\sigma_0^2 +(1-w)^2n_r n_e^{-1} \sigma_e^2}} \leq \frac{ \theta_0 - \theta^\star }{  \sqrt{\pi_1^{-1} \sigma_1^2 + \pi_0^{-1} \sigma_0^2 }}. 
\end{align*}
For instance, when $\Delta_0 = \Delta^\star $, i.e., the specified upper bound for $\Delta^\star $ is tight, and $\sigma_{0}^2 = \sigma_{e}^2$, i.e., the variance of $Y$ for the two control groups are equal, simple algebra reveals that the power of $T_{2, \Delta_0} (w)$ is always larger than that of $T_1$ for any $w $ satisfying $ {\color{black} \max(0, (n_r\pi_0  - n_e )/(n_r\pi_0 + n_e) ) } \leq w<1$. 

Second, we can derive the oracle $w$ that maximizes the power of $T_{2, \Delta_0} (w)$. Let  $\kappa=(\pi_0^{-1}\sigma_0^2)/(\pi_1^{-1}\sigma_1^2 + \pi_0^{-1}\sigma_0^2)$, the optimal $w$ takes the following form:
\begin{align}
	w_{\rm opt} = \left\{ 
	\begin{array}{cc}
		1, &  \text{when } \Delta_0-\Delta^\star \textcolor{black}{\geq} \kappa(\theta^\star -\theta_0)>0, \\
		1-\frac{(\Delta_0- \Delta^\star)  (\pi_1^{-1}\sigma_1^2 + \pi_0^{-1}\sigma_0^2) + (\theta_0 - \theta^\star) \pi_0^{-1}\sigma_0^2  }{ (\theta_0 - \theta^\star) (n_r n_e^{-1}\sigma_e^2 +\pi_0^{-1}\sigma_0^2)   + (\Delta_0- \Delta^\star) \pi_0^{-1}\sigma_0^2 },  &  \text{when }  \kappa(\theta^\star -\theta_0)> \Delta_0-\Delta^\star \textcolor{black}{\geq} 0 ,
	\end{array}
	\right.\label{eq: w opt}
\end{align}
where the first case is
when $\Delta_0$ is specified too large,  the power of $ T_{2, \Delta_0}(w)$ is maximized at $w=1$, which means that using the external controls does not  lead to efficiency gain. 
As an illustration, under the special case that  $\pi_1^{-1}\sigma_1^2= \pi_0^{-1}\sigma_0^2 = n_r n_e^{-1}\sigma_e^2$, when $\Delta_0- \Delta^\star >(\theta^\star - \theta_0) /2$, the optimal $w$ is 1, whereas when $(\theta^\star - \theta_0) /2 > \Delta_0- \Delta^\star > 0$, the optimal $w$ is $1-\{ (\theta_0 - \theta^\star ) +2 (\Delta_0 - \Delta^\star)\}/\{2(\theta_0 - \theta^\star ) + (\Delta_0 - \Delta^\star)\}$. Under another special case when $\Delta^\star = \Delta_0$ and $\sigma_1= \sigma_0= \sigma_e$, $w_{\rm opt}$ becomes $(n_r\pi_0 ) /(n_r\pi_0 + n_e)$, which agrees with the optimal $w$ under exchangeability discussed in Section \ref{subsec: testing Exchangeability}. The proof of (\ref{eq: w opt}) is given in the supplementary material.

Lastly, we compare the two tests $T_1$ and $T_{2, \Delta_0}(w)$ in terms of their limiting power as the sample sizes grow to infinity. When $\theta^\star >\theta_0$ and $\lim\limits_{n_r\rightarrow +\infty} \sqrt{n_r}(\theta^\star-\theta_0)= +\infty$ (e.g., when $\theta^\star, \theta_0$ are two constants), then  the power of $T_1$ goes to 1 as $n_r\rightarrow \infty$. In contrast, the limiting power of $ T_{2, \Delta_0}(w)$ depends on specifications of $w$ and $\Delta_0$. In particular, there exists a $w$-dependent number $\widetilde \Delta (w) =(\theta^\star  - \theta_0 ) /(1-w)+ \Delta^\star $, such that the power of $ T_{2, \Delta_0}(w)$ tends to 1 if $\Delta_0 <  \widetilde \Delta (w)$ and to 0 if $\Delta_0 > \widetilde \Delta (w)$ as $\min(n_r, n_e) \rightarrow + \infty$, 
so $  \widetilde \Delta (w)$ characterizes the limiting behavior of $ T_{2, \Delta_0}(w)$ under the alternative. This number  $\widetilde \Delta (w) $ is analogous to the design sensitivity in the literature of sensitivity analysis \citep{rosenbaum2004design, rosenbaum2020design}.



\section{A Combined Test}
\label{sec: combined test}

{\color{black} Should we leverage external controls? In other words, is it better to use the test statistic $T_1$ constructed solely based on the RCT or the test statistic $ T_{2, \Delta_0}(w)$ that additionally leverages the external controls?} We know from the above theory and analysis that the answer to this question depends upon the context, specifically upon the nature and size of the treatment effect, and the specification of $w$ and $\Delta_0$, that might be difficult to anticipate prior to examining the data. {\color{black} As Motivated in Section \ref{sec: intro}, we propose a combined testing procedure that performs both $T_1$ and $ T_{2, \Delta_0}(w)$, correcting for multiple testing using the joint distribution of the two test statistics.}

Under $H_0$, the joint distribution of $(T_1, T_{2,\Delta^\star} (w) )$ is asymptotically bivariate normal, satisfying 
$$
\left(
\begin{array}{c}
	T_1  \\
	T_{2,\Delta^\star} (w)
\end{array}\right) 
\xrightarrow{d} N\left(
\left[\begin{array}{cc}
	0 \\0 \end{array}\right], 
\left[\begin{array}{cc}
	1 & \rho\\
	\rho & 1
\end{array}\right]\right),
$$
where 
$$
\textcolor{black}{\rho =\frac{\pi_1^{-1}\sigma_1^2 + w\pi_0^{-1}\sigma_0^2}
	{\sqrt{(\pi_1^{-1}\sigma_1^2 + \pi_0^{-1}\sigma_0^2)(\pi_1^{-1}\sigma_1^2 +w^2 \pi_0^{-1}\sigma_0^2+(1-w)^2 n_r n_e^{-1}\sigma_e^2)}}}.
$$
Again, for illustration, consider the special case that   $\pi_1^{-1}\sigma_1^2= \pi_0^{-1}\sigma_0^2 = n_r n_e^{-1}\sigma_e^2$, then $\rho$ increases as $w$ increases from 0 to 1, and thus $\rho$ ranges between 0.5 and 1. 

Consider the testing procedure that, for any specified $\Delta_0 $ and $w$, rejects $H_0$ if 
\begin{align}
	\max(T_1, T_{2, \Delta_0} (w) ) \geq c_{1-\alpha; \rho}, \label{eq: rej combine}
\end{align}
where $c_{1-\alpha; \rho}$ satisfies $\Phi_{2,\rho} ( c_{1-\alpha; \rho}) = 1- \alpha $, $\Phi_{2,\rho} (x,y)$ is the probability of the 2-dimensional lower orthant $(-\infty, x] \times (-\infty, y] $ for a bivariate normal distribution with expectation $(0,0)^T$, unit variances, and correlation coefficient $\rho$, and write $\Phi_{2,\rho} (x)=\Phi_{2,\rho} (x,x) $.  This combined testing procedure is able to control the type I error for any $\Delta^\star\in [-\infty, \Delta_0] $ because 
$$
P_{H_0}\left(\max(T_1, T_{2, \Delta_0}(w) )\geq c_{1-\alpha; \rho} \right)
\leq 
P_{H_0, \Delta^\star =\Delta_0}(\max(T_1, T_{2, {\Delta^\star}}(w))\geq c_{1-\alpha; \rho}) = \alpha.
$$

In what follows, we establish several attractive features of the combined test. Note that 
under the alternative hypothesis, the power of the combined test -- the probability of event  \eqref{eq: rej combine} --  is 
\begin{align}
	& P(\max(T_1, T_{2,\Delta_0} (w) )\geq c_{1-\alpha; \rho})  \nonumber \\
	&\approx 1-\Phi_{2,\rho}\left(
	c_{1-\alpha; \rho} + \underbrace{\frac{ \sqrt{n_r} (\theta_0 - \theta^\star ) }{ \sqrt{ \pi_1^{-1} \sigma_1^2 + \pi_0^{-1} \sigma_0^2}  }}_{B_1} ,  \ c_{1-\alpha; \rho} + \underbrace{ \frac{\sqrt{n_r} (\theta_0 - \theta^\star )+  \sqrt{n_r} (1-w)(\Delta_0-\Delta^\star) }{\sqrt{ \pi_1^{-1}\sigma_1^2 +w^2  \pi_0^{-1}\sigma_0^2 +(1-w)^2n_r n_e^{-1} \sigma_e^2}}}_{B_2}\right), \label{eq: powerc}
\end{align}
where $\approx$ means asymptotic approximation. This leads to the first observation that the power of the combined test is generally larger than the worst of the two component tests, i.e., $\text{Power}_c  \geq \min (\text{Power}_1, \text{Power}_2 )$, where $\text{Power}_1, \text{Power}_2, \text{Power}_c$ are respectively the asymptotic power of $T_1, T_{2, \Delta_0}(w),$ and the combined test. This can be seen from noting that 
\begin{align*}
	1-\text{Power}_c &= \Phi_{2, \rho}(c_{1-\alpha; \rho} + B_1 , c_{1-\alpha; \rho}+ B_2) \\
	&\leq \Phi_{2, \rho}(c_{1-\alpha; \rho}+ \min (B_1,B_2), +\infty)\\
	&=\Phi(c_{1-\alpha; \rho}+ \min (B_1,B_2))\\
	&=\Phi(z_{1-\alpha}+ \max (B_1,B_2)- \{|B_1-B_2|-(c_{1-\alpha; \rho}-z_{1-\alpha})\})\\
	&\leq \Phi(z_{1-\alpha}+ \max(B_1,B_2) )\\
	&= \max \left\{ \Phi(z_{1-\alpha}+ B_1 ),  \Phi(z_{1-\alpha}+ B_2 ) \right\}\\
	&= 1-  \min (\text{Power}_1, \text{Power}_2 ),
\end{align*}
where the  second inequality holds when {\color{black} $|B_1-B_2|\geq(c_{1-\alpha; \rho}-z_{1-\alpha})$}, i.e., when the power of the two component tests are not too similar. 

Moreover, not only is the power of the combined test better than the worst of the two  component tests in finite sample, it is also close to the better of the two component tests in finite sample, and equal to the better of the two component tests in the limit. To see this, we bound the difference in power as follows 
\begin{align}
	\max(\Power_1,\Power_2 ) - \Power_c &= \Phi_{2, \rho}( c_{1-\alpha; \rho}+ B_1, c_{1-\alpha; \rho}+ B_2) -\Phi(z_{1-\alpha}+ \min  (B_1,B_2) )\nonumber \\
	&\leq \Phi(c_{1-\alpha; \rho}+ \min  (B_1,B_2))-\Phi( z_{1-\alpha}+ \min  (B_1,B_2)) \nonumber\\
	&\leq 1-2\Phi((z_{1-\alpha}-c_{1-\alpha; \rho})/2). \nonumber
\end{align}
It is helpful to anchor several values of $c_{1-\alpha; \rho}$ and the upper bound $1-2\Phi((z_{1-\alpha}-c_{1-\alpha; \rho})/2)$  in terms of different $\alpha$ and $\rho$. When  $\alpha = 0.025$ and for $\rho=0.5, 0.7, 1$, the critical values are 
$c_{1-\alpha;0.5} = 2.21, c_{1-\alpha;0.7} = 2.18, c_{1-\alpha;1} = 1.96$, and  correspondingly, the upper bounds are 0.100, 0.088, 0.   This means that because of the high correlation between $T_1$ and $ T_{2,\Delta^\star} (w)$, the price paid for multiple testing is generally small.  With regard to the limiting power, it is also easy to see that for fixed $\theta_0$ and $\theta^\star>\theta_0$, $B_1\rightarrow - \infty$ as the sample size $n_r$ increases. Hence, the combined test always has its power approaching 1 as $n_r\rightarrow \infty$, just like the test $T_1$ that only uses RCT data, {\color{black} which is not the case for $ T_{2,\Delta^\star} (w)$ as discussed in Section \ref{subsec: power comparison without exchangeability}. This further shows the advantage of the combined test.}

{\color{black} For implementation of the sensitivity analysis (either $T_{2, \Delta_0}$ or the combined test), practitioners are not required to specify the value of the sensitivity parameter $\Delta_0$. Following the pioneering work by \citet{Cornfield1959} and the sensitivity analysis literature \cite{rosenbaum2020design}, results from the combined test can be summarized by the ``tipping point'' -- the magnitude of $\Delta_0$ that would be needed such that the null hypothesis can no longer be rejected. If such a value of $\Delta_0$ is deemed implausible, then we still have evidence to reject the null hypothesis based on the combined test. In Section \ref{sec: real}, we illustrate the method using a real example. 
}

\section{Power Calculations}
\label{sec: simu}
We investigate three factors when conducting power calculations. The first factor concerns the true treatment effect $\theta^\star = 0.2, 0.3,$ and 0.4. The second factor is the specified value of the maximum bias $\Delta_0 = 0.2, 0.3,  0.4, 0.6$. The third factor is the sample size $n_1 = 50, 100, 150, 200$, with $n_1: n_0:n_e= 2:1:3$.  Additional parameters are $\theta_0 =0$,  $\Delta^\star = 0.2$,  $ \sigma_1= \sigma_0 = \sigma_e = 1$, and $\alpha=0.025$. 

Table \ref{tab:simu formula power} summarizes the power of  $T_1,  T_{2, \Delta_0}(w)$ and the combined test  $T_{c, \Delta_0}(w)= \max(T_1, T_{2, \Delta_0}(w))$, calculated respectively using \eqref{eq: power1}, \eqref{eq: power2}, and \eqref{eq: powerc}.  For $ T_{2, \Delta_0}(w)$ and $T_{c, \Delta_0}(w)$, we consider two choices of $w$: the oracle $w$ in \eqref{eq: w opt} that maximizes the power (denoted as $w_{\rm opt}$), and its value under exchangeability $n_0/(n_0+ n_e)= 1/4$.  In the supplementary material, we check powers by simulation, finding good agreement. {\color{black} In the supplementary material, we also include a check of the type I error, which are all close to or below the nominal level, indicating validity of all the tests. In contrast, a naive combined test without correcting for multiple testing cannot control the type I error.
}

{\color{black}The following is a summary of results in Table \ref{tab:simu formula power}. 
	\begin{enumerate}
		\item Across all scenarios, the power of the combined test $T_{c, \Delta_0}(1/4)$ is larger than the worst of the power of $T_1$ and $T_{2, \Delta_0}(1/4)$,  and close to the best of the power of $T_1$ and $T_{2, \Delta_0}(1/4)$. This supports our theory in Section \ref{sec: combined test}. 
		\item For $T_1$, its power is not affected by $\Delta_0$.  For $T_{2, \Delta_0}(1/4)$,  
		its power is mostly larger than that of $T_1$ when $\Delta_0= 0.2, 0.3$, but quickly diminishes as $\Delta_0$ increases and becomes 
		substantially smaller than that of  $T_1$ when $\Delta_0= 0.4, 0.6$ across most scenarios.   In comparison,  when $\theta^\star= 0.2, 0.3$, the sensitivity parameter $\Delta_0$ can be as large as 0.3 before the combined test $T_{c, \Delta_0}(1/4)$ starts to lose power compared to $T_1$; when $\theta^\star=0.4$,   the sensitivity parameter $\Delta_0$ can be as large as 0.4. If a $\Delta_0$ larger than 0.3 or 0.4 is deemed implausible by practitioners, the combined test  $T_{c, \Delta_0}(1/4)$ will have power gain compared to $T_1$. 
		On the other hand, because the combined test $T_{c, \Delta_0}(1/4)$ still performs $T_1$ as one of its component (i.e., anchors at $T_1$)  but with a small adjustment for testing twice, the potential power loss compared to $T_1$ is never too large. This clearly demonstrates the key advantage of the combined test.
		\item As the sample size increases, the power of $T_1$ and $T_{c, \Delta_0}(1/4)$ always increases. However, as the sample size increases, the power of $T_{2, \Delta_0}(1/4)$ tends to 0 when $\theta^\star =0.2$ and $\Delta_0=0.6$,  stays unchanged when $\theta^\star =0.3$ and $\Delta_0=0.6$, and tends to 1 in other cases. This behavior of $T_{2, \Delta_0}(1/4)$  supports the result that  the power of  $T_{2, \Delta_0}(1/4)$ tends to 1 when $\Delta_0 < \tilde\Delta(1/4)$ and to 0 when $\Delta_0 > \tilde\Delta(1/4)$ as the sample size increases, where  $\tilde\Delta(1/4) $ defined in Section \ref{subsec: power comparison without exchangeability} equals $ 4\theta^\star/3 + 0.2$, which is 0.47, 0.60, and 0.73 for $\theta^\star = 0.2, 0.3, 0.4$, respectively.
		\item 
		{Lastly, the oracle tests $T_{2, \Delta_0}(w_{\rm opt})$ and $T_{c, \Delta_0}(w_{\rm opt})$ are included as a reference. The test $T_{2, \Delta_0}(w_{\rm opt})$ is more powerful than $T_1$ and $T_{2, \Delta_0}(1/4)$, which agrees with our theory as $T_{2, \Delta_0}(w_{\rm opt})$ maximizes power among a family of test statistics $\{T_2(w), w\in [0,1]\}$. Observing that the power of  $T_{c, \Delta_0}(1/4)$ and $T_{c, \Delta_0}(w_{\rm opt})$ are similar indicates that setting $w=n_0/(n_0+ n_e) $ usually leads to desirable power performance.}  
\end{enumerate}}

\section{Application}
\label{sec: real}

{\color{black} We revisit the example introduced in Section \ref{subsec: motivating example} and illustrate how the proposed methods can be applied.}
Formally, we test the hypothesis that $H_0: \theta^\star = \theta_0$ versus $H_A: \theta^\star < \theta_0$, with $\theta_0 = 0.4$, which  can be equivalently implemented using the tests described in Sections \ref{sec: testing}-\ref{sec: combined test} with $Y_i$'s replaced by $- Y_i$'s and $\theta_0$ replaced by $-\theta_0$. We set the significance level $\alpha=0.025$.

Using only the internal RCT,  $T_1=4.80$ with p-value $7.92\times 10^{-7}$, based on which we reject the null hypothesis $H_0$. This result is solely based on internal controls and thus is invariant to the value of $\Delta_0$. 

{\color{black} Leveraging external controls and let $w=n_0/(n_0+n_e) = 0.485$,  $T_{2} (w) = 5.08  $ with p-value $1.88\times 10^{-7}$ when $\Delta_0 =0$. Therefore, under the exchangeability assumption, we can also reject the null hypothesis $H_0$. To gauge the robustness of this conclusion to violation of the exchangeability, we apply the proposed sensitivity analysis. As discussed at the end of Section \ref{sec: combined test}, results of our sensitivity analysis can be summarized by the ``tipping point'' - the magnitude of $\Delta_0$ that would be needed such that the null hypothesis can no longer be rejected. In this example, as $\Delta_0$ increases, the adjusted p-value associated with $T_{2, \Delta_0} (w)$ increases but remains below $\alpha=0.025$ for any $\Delta_0 \leq 0.62$. Namely, two patients with the same observed characteristics (as listed in Table \ref{tb: table1}), one in the internal RCT and the other in the external trial, may differ in their expected potential outcome under control by up to 0.62, under which the adjusted p-value is still below the significance level $\alpha$. This means that the significant effect we observe cannot be explained away by unmeasured biases of magnitude up to $\Delta_0=0.62$. If such a large unmeasured bias is deemed implausible, then there is no real doubt that the rejection based on $T_{2, \Delta_0}$ provides evidence of noninferiority.} 

Finally, using the combined test,  $\max (T_1,  T_{2, \Delta_0}(w))= 5.08$ with adjusted p-value $3.41\times 10^{-7}$ when $\Delta_0=0$.  As $\Delta_0$ increases, the adjusted p-value for the combined test increases but plateaus at $1.41\times 10^{-6}$ when $T_1\geq  T_{2, \Delta_0}(w)$. This means that rejection based on the combined test is insensitive to any value of $\Delta_0$, i.e.,  similar to $T_1$ that only uses the internal RCT, rejection based on the combined test is insensitive to any violation of the exchangeability assumption. 



{\color{black} 
	It is also interesting to see the relative performance of $T_1,  T_{2, \Delta_0}(w),  T_{c, \Delta_0}(w)$ when the internal RCT is underpowered, and thus the combined test may be more useful. For this purpose, we randomly sample with replacement 100 patients from the internal RCT, with a target ratio of 4/5 from the treated arm and 1/5 from the control arm. Then $T_1$ is computed using this subsample from the RCT, while $T_{2, \Delta_0}(w)$ and $ T_{c, \Delta_0}(w)$ additionally use the external controls that were matched to the sampled treated patients with  $w=n_0/(n_0+n_e)$ calculated using the subsample. This procedure is repeated 1000 times. Among these repetitions, $T_1$  rejects the null hypothesis 71.5\% of the time, i.e., the power of $T_1$ is 71.5\%, while the combined test $T_{c, \Delta_0}(w)$ has power 82.4\%, 74.2\%,  71.1\% when $\Delta_0 = 0.1, 0.2, 0.25$, respectively.  Hence, the sensitivity parameter $\Delta_0$ can be as large as 0.25 before the combined test starts to lose power compared to $T_1$. In comparison,  $T_{2, \Delta_0}(w)$ has worse performance, with power equal to 80.4\%, 64.8\%,  45.7\% when $\Delta_0 = 0.1, 0.2, 0.25$, respectively. Taking a closer look at the results, we note that if $T_1$ is larger than $ c_{1-\alpha; \rho}$ defined in \eqref{eq: rej combine}, then both $T_1$ and the combined test  can reject  $H_0$  regardless of the value of $\Delta_0$. If $T_1< z_{1-\alpha}$, then $T_1$ cannot reject $H_0$ while the combined test can still reject 27.7\% of these cases at $\Delta_0=0.2$. The potential loss of using the combined test is when $T_1$ is between $z_{1-\alpha}$ and $ c_{1-\alpha; \rho}$, in which cases using $T_1$ alone can reject $H_0$ but  the combined test is sensitive to a certain value of $\Delta_0$. However, this scenario is relatively rare and occurs in 8.4\% of the repetitions; furthermore, even in this scenario, the combined test can still reject $H_0$ at $\Delta_0=0.2$ around half the time. 
	
	The last step of a sensitivity analysis is to reason about whether a value of $\Delta_0=0.2$ is plausible given that we have already controlled for baseline covariates listed in Table \ref{tb: table1}. For this task, an intuitive strategy is to judge the plausibility of $\Delta_0$ in reference to some observed covariates \citep{Imbens:2003}. Specifically, we can omit observed covariates one at a time during matching and calculate $\bar Y_0- \bar Y_e$ using the resulting matched external controls. Using this procedure, we estimate the amount of bias from not matching on one of the observed covariates and to benchmark the plausibility of $\Delta_0$, the amount of bias from not being able to match on the region variable. The results show that omitting the baseline HbA1c leads to the largest $\bar Y_0- \bar Y_e$ that is equal to  0.14, while omitting any other observed variables in Table \ref{tb: table1} leads to   $\bar Y_0- \bar Y_e$ ranging from -0.05 to 0.04.  
	Based on the prior knowledge in \citet{home2014predictive} that the baseline HbA1c explains most of the variability in the change in HbA1c, particularly in comparison to the geographical region, we view that $\Delta=0.2$ is implausible. 
}

{\color{black} 
	In summary, before looking at the data, the choice between $T_1$ and $ T_{2, \Delta_0}(w)$, would be difficult to make or justify on the basis of a priori considerations. In some cases, $T_1$ may not be powerful enough due to the small sample size of the internal RCT, while leveraging external controls leads to a more powerful test. In some other cases, $ T_{2, \Delta_0}(w)$ may be sensitive to unmeasured biases while $T_1$ is already powerful enough. Under these circumstances, the combined test $ T_{c, \Delta_0}(w)$ is often preferable as it performs both tests with a small correction for multiple testing by taking into account the high correlation of the two test statistics.}

\section{Discussion}
\label{sec: discuss}

We propose a sensitivity analysis approach for using external controls in clinical trials to examine the robustness of study conclusion to remaining unmeasured bias after controlling for measured covariates. {\color{black} Results from the sensitivity analysis can be summarized by the ``tipping point'' -- the magnitude of $\Delta_0$ that would be needed such that the null hypothesis can no longer be rejected. If $\Delta_0$ is deemed plausible (or implausible), the conclusion based on using external controls is sensitive (or robust) to unmeasured bias.}

When in doubt about whether the use of external controls increases power, 
we propose a combined testing procedure that performs both tests, one only using the internal controls and one additionally using the external controls, correcting for multiple testing using the joint distribution of the two test statistics. {\color{black} Because the two test statistics are highly correlated, this correction for multiple testing is small, and thus the combined test only has a small loss of power compared to knowing a priori which test is best. Moreover, the combined test provides a new method of sensitivity analysis designed for data fusion problems, which 
	anchors at the unbiased RCT-only analysis and spends a small proportion of the type I error to also test using the external controls. In this way, if leveraging external controls increases power, the power gain compared to the RCT-only analysis   can be substantial; if not, the power loss is small.
}



Our work is motivated by the literature of sensitivity analysis, in which testing a hypothesis multiple times has been shown to be useful {\color{black} in enhancing the robustness to unmeasured bias}
\citep{rosenbaum2012testing,Small:2013aa, Rosenbaum:2017aa, ye2021combining}. {\color{black} Nonetheless, we focus on a distinct context and have shown that testing multiple times using both a known unbiased test and potentially biased tests can be particularly attractive for data fusion problems. We also have  developed various properties of the combined procedure that has not appeared in the existing literature.}

Finally, a remaining question is how to choose $w$ for the combined test. The power of the combined test depends on $w$ in a complicated way as $w$ not only affects the definition of $T_{2, \Delta_0} (w)$ but also the correlation $\rho$, which makes finding the optimal $w$ a cumbersome task. In practice, a reasonable choice is $w= \pi_0 n_r/(n_e+ \pi_0 n_r) $, which minimizes the variance of $w\overline{Y}_0+ (1-w)\overline{Y}_e$ when $ \Var(Y^{(0)}\mid D=1) = \Var(Y^{(0)}\mid D=0)  $. Another way is to pre-specify several values of $w$, calculate the corresponding test statistics, and combine all the test statistics using their joint null distribution. Because of the high correlation between these test statistics, the price paid for multiple testing will generally be small.

\bibliographystyle{apalike}
\bibliography{combine}

\begin{table}[ht]
	\centering
	\resizebox{0.88\textwidth}{!}{
		\begin{threeparttable}
			\caption{ Covariate balance after matching in 159 matched pairs of one treated patient in the RCT and one external control patient.\label{tb: table1}}
			\begin{tabular}{lccc}
				& Treated & External Control &  Standardized \\ 
				&   ($ n_1= $159)& ($ n_e= $159) & Mean Difference\\   
				\hline
				Age (years)   & 57.45 & 57.16 & 0.03 \\ 
				Female (fr) & 0.41 & 0.41 & 0.00 \\
				Overweight (fr) & 0.28 & 0.28 & 0.00 \\ 
				Obese (fr) & 0.72 & 0.72 & 0.00 \\ 
				Diabetes Duration (years) & 12.08 & 11.74 & 0.05 \\ 
				Hypertension (fr) & 0.66 & 0.69 & -0.07 \\ 
				History of MI (fr) & 0.04 & 0.01 & 0.13 \\ 
				History of CR (fr) & 0.04 & 0.02 & 0.13 \\ 
				History of CABG (fr) & 0.01 & 0.00 & 0.05 \\ 
				Lipid Lowering Medication (fr) & 0.61 & 0.57 & 0.09 \\ 
				Statin Use (fr) & 0.51 & 0.50 & 0.01 \\ 
				Non-Statin Lipid Lowering Medication (fr) & 0.15 & 0.11 & 0.10 \\ 
				Fasting Serum Glucose (mg/dL) & 164.99 & 166.25 & -0.03 \\ 
				Triglycerides (mg/dL) & 139.86 & 140.06 & -0.00 \\ 
				Total Cholesterol (mg/dL) & 178.92 & 180.93 & -0.05 \\ 
				LDL (mg/dL) & 101.42 & 103.29 & -0.06 \\ 
				HDL (mg/dL) & 50.41 & 50.01 & 0.03 \\ 
				Alanine Aminotransferase (U/L) & 33.60 & 33.03 & 0.03 \\ 
				Aspartate Aminotransferase (U/L) & 26.99 & 26.08 & 0.08 \\ 
				Total Bilirubin (mg/dL) & 0.58 & 0.54 & 0.12 \\ 
				eGFR (mL/min/1.73m$^2$) & 90.52 & 87.70 & 0.13 \\ 
				Baseline Sulfonylureas or Meglitinides Use (fr) & 0.86 & 0.86 & 0.02 \\ 
				Smoking (fr) & 0.46 & 0.43 & 0.06 \\ 
				Baseline HbA1c (\%) & 8.57 & 8.59 & -0.03 \\ 
			\end{tabular}
			\begin{tablenotes}
				\small
				\item Abbreviations:  CABG = coronary artery bypass
				graft; CR = coronary revascularization; eGFR = estimated glomerular filtration rate based on the modified Modification of Diet in
				Renal Disease equation; fr = fraction; HbA1c = hemoglobin A1c; HDL = high-density lipoprotein cholesterol; LDL =
				low-density lipoprotein cholesterol; MI = myocardial infarction. 
			\end{tablenotes}
	\end{threeparttable}}
\end{table}

\renewcommand{\baselinestretch}{1.1}
\hspace{-2cm}
\begin{table}[h!tb] \centering
	\caption{Theoretical power (in \%) for $T_1,  T_{2, \Delta_0}(w)$ and the combined test $T_{c, \Delta_0} (w)$ with $w=1/4 $ or $  w_{\rm opt}$, where $\theta_0 = 0$, $\Delta^\star = 0.2$, $n_{1}: n_{0}: n_{e} = 2:1:3$,  $\sigma_1 = \sigma_0 = \sigma_{e} = 1$, and $\alpha=2.5\%$. In the table, we omit the $\Delta_0$ subscript for notational simplicity.
	}
	\label{tab:simu formula power}\vspace{3mm}
	\tabcolsep= 3pt 
	\resizebox{\textwidth}{!}{\begin{tabular}{ccccccccccccccccccccccccccc}
			& &
			\multicolumn{5}{c}{$\theta^\star = 0.2$}   & &
			\multicolumn{5}{c}{$\theta^\star = 0.3$} & &
			\multicolumn{5}{c}{$\theta^\star = 0.4$}\\
			\cmidrule{3-7} \cmidrule{9-13} \cmidrule{15-19}
			$\Delta_0$ &$n_1$ &
			$T_1$&
			$T_{2}(1/4)$ &  
			$T_{2}(w_{\rm opt})$ &     
			$T_{c}(1/4)$&   
			$T_{c}(w_{\rm opt})$ & &
			$T_1$&
			$T_{2}(1/4)$ & 
			$T_{2}(w_{\rm opt})$ &     
			$T_c(1/4)$&   
			$T_c(w_{\rm opt})$ & &
			$T_1$&
			$T_{2}(1/4)$ & 
			$T_{2}(w_{\rm opt})$ &     
			$T_c(1/4)$&   
			$T_c(w_{\rm opt})$\\     
			\hline
			\multirow{4}{*}{0.2} & 50  & 12.6 & 21.0 & 21.0 & 18.5 & 18.5 &  & 23.1 & 41.0 & 41.0 & 36.5 & 36.5 &  & 37.2 & 63.7 & 63.7 & 58.4 & 58.4 \\
			& 100 & 21.0 & 37.2 & 37.2 & 33.0 & 33.0 &  & 41.0 & 68.8 & 68.8 & 63.7 & 63.7 &  & 63.7 & 90.4 & 90.4 & 87.5 & 87.5 \\
			& 150 & 29.3 & 51.6 & 51.6 & 46.5 & 46.5 &  & 56.4 & 85.1 & 85.1 & 81.3 & 81.3 &  & 80.7 & 97.9 & 97.9 & 97.0 & 97.0 \\
			& 200 & 37.2 & 63.7 & 63.7 & 58.4 & 58.4 &  & 68.8 & 93.4 & 93.4 & 91.1 & 91.1 &  & 90.4 & 99.6 & 99.6 & 99.4 & 99.4 \\
			\hline
			\multirow{4}{*}{0.3} & 50  & 12.6 & 10.8 & 13.2 & 12.4 & 13.0   &  & 23.1 & 25.4 & 27.7 & 26.1 & 26.4 &  & 37.2 & 46.7 & 48.6 & 45.6 & 45.7 \\
			& 100 & 21.0   & 17.4 & 22.1 & 20.6 & 21.7 &  & 41.0   & 45.1 & 49.1 & 46.3 & 46.9 &  & 63.7 & 75.6 & 77.7 & 74.7 & 74.9 \\
			& 150 & 29.3 & 23.9 & 30.8 & 28.7 & 30.3 &  & 56.4 & 61.4 & 66.0   & 63.0   & 63.7 &  & 80.7 & 90.1 & 91.6 & 89.7 & 89.9 \\
			& 200 & 37.2 & 30.3 & 39.1 & 36.6 & 38.5 &  & 68.8 & 73.8 & 78.2 & 75.4 & 76.1 &  & 90.4 & 96.3 & 97.1 & 96.2 & 96.3 \\
			\hline
			\multirow{4}{*}{0.4} & 50  & 12.6 & 4.7 & 12.6 & 9.8 & 12.6 &  & 23.1 & 13.7 & 23.1 & 20.3 & 23.1 &  & 37.2 & 30.3 & 39.1 & 36.6 & 38.5 \\
			& 100 & 21.0 & 6.0 & 21.0 & 16.3 & 21.0 &  & 41.0 & 23.1 & 41.0 & 36.5 & 41.0 &  & 63.7 & 53.2 & 66.3 & 63.1 & 65.5 \\
			& 150 & 29.3 & 7.2 & 29.3 & 23.0 & 29.3 &  & 56.4 & 32.3 & 56.4 & 51.2 & 56.4 &  & 80.7 & 70.5 & 83.0 & 80.4 & 82.4 \\
			& 200 & 37.2 & 8.3 & 37.2 & 29.9 & 37.2 &  & 68.8 & 41.0 & 68.8 & 63.7 & 68.8 &  & 90.4 & 82.3 & 92.0 & 90.3 & 91.6 \\
			\hline                     
			\multirow{4}{*}{0.6} & 50  & 12.6 & 0.6 & 12.6 & 8.7 & 12.6 &  & 23.1 & 2.5 & 23.1 & 17.2 & 23.1 &  & 37.2 & 8.3 & 37.2 & 29.9 & 37.2 \\
			& 100 & 21.0 & 0.3 & 21.0 & 15.3 & 21.0 &  & 41.0 & 2.5 & 41.0 & 32.8 & 41.0 &  & 63.7 & 12.6 & 63.7 & 55.5 & 63.7 \\
			& 150 & 29.3 & 0.2 & 29.3 & 22.2 & 29.3 &  & 56.4 & 2.5 & 56.4 & 47.7 & 56.4 &  & 80.7 & 16.9 & 80.7 & 74.3 & 80.7 \\
			& 200 & 37.2 & 0.1 & 37.2 & 29.3 & 37.2 &  & 68.8 & 2.5 & 68.8 & 60.7 & 68.8 &  & 90.4 & 21.0 & 90.4 & 86.2 & 90.4\\
	\end{tabular}}
\end{table}

\clearpage 

\setcounter{equation}{0}
\setcounter{table}{0}
\setcounter{section}{0}
\setcounter{figure}{0}
\renewcommand{\theequation}{S\arabic{equation}}
\renewcommand{\thefigure}{S\arabic{figure}}
\renewcommand{\thetable}{S\arabic{table}}

\begin{center}
	{\sffamily\bfseries\LARGE
		Supplementary Materials
	}
\end{center}
\section{Additional type I error and power calculations}

We conduct the type I error calculation in two approaches. 
For $T_1,  T_{2, \Delta_0}(w)$ and the combined test  $T_{c, \Delta_0}(w)$, the theoretical version is calculated using formulas in \eqref{eq: power1}, \eqref{eq: power2}, and \eqref{eq: powerc}, respectively; the empirical version is obtained from 10,000 simulation repetitions. {\color{black}  For comparison, we also include a naive combined test $\widetilde T_{c, \Delta_0} (w)$ that performs both $T_1$ and $T_{2, \Delta_0}(w)$ without adjusting for multiple testing, i.e., it rejects $H_0$ if $\max(T_1, T_{2, \Delta_0} (w) ) \geq z_{1-\alpha}$.} In each repetition, $n_1$ treated subjects are generated from $Y^{(1)}\mid D=1\sim N(0, \sigma_1^2) $, $n_0$ internal controls from $Y^{(0)}\mid D=1 \sim N( -\theta^\star, \sigma_0^2) $, and $n_e$ external controls from $Y^{(0)}\mid D=0 \sim N( -\theta^\star - \Delta^\star , \sigma_e^2) $. The results  are in Table \ref{tab:simu type I}. From Table \ref{tab:simu type I}, we see that the type I error rates of $T_1, T_2(1/4), T_c(1/4)$ are close to or below the significance level. However, using the naive combined test $\widetilde T_{c, \Delta_0} (w)$ can lead to inflated type I error. 

The power calculation is performed similarly; the only difference is that the empirical power is obtained from 3,000 simulation repetitions. The results are in Table \ref{tab:simu empirical power}.

\renewcommand{\baselinestretch}{1.1}
\hspace{-2cm}
\begin{table}[h!tb] \centering
\caption{ Empirical and theoretical type I error (in \%) for $T_1,  T_{2, \Delta_0}(w)$, the combined test $T_{c, \Delta_0} (w)$, and the naive combined test $\widetilde T_{c, \Delta_0} (w)$  with $w=1/4 $, where $\theta_0 = 0$, $\Delta^\star = 0.2$, $n_{1}: n_{0}: n_{e} = 2:1:3$,  $\sigma_1 = \sigma_0 = \sigma_{e} = 1$, and $\alpha=2.5\%$. The empirical version is based on 10,000 repetitions. In the table, we omit the $\Delta_0$ subscript for notational simplicity.}
\vspace{3mm}
\label{tab:simu type I}
\tabcolsep= 3pt 
\resizebox{!}{!}{
\begin{tabular}{cccccccccccccccccccccccccc}
	& &
	\multicolumn{4}{c}{Empirical Type I error}   & &
	\multicolumn{4}{c}{Theoretical Type I error}\\
	\cmidrule{3-6} \cmidrule{8-11} 
	$\Delta_0$ &$n_1$ &
	$T_1$&
	$T_{2}(1/4)$ &  
	$T_c(1/4)$&   
	$\widetilde T_c(1/4)$&   
	&
	$T_1$&
	$T_{2}(1/4)$ & 
	$T_c(1/4)$&   
	$\widetilde T_c(1/4)$ \\ 
	\hline
	\multirow{4}{*}{0.2} 
	& 50  & 2.9 & 2.6 & 2.8 &4.6 &  & 2.5 & 2.5 & 2.5 &4.2\\
	& 100 & 2.7 & 2.4 & 2.7 &4.4 &  & 2.5 & 2.5 & 2.5 &4.2\\
	& 150 & 2.6 & 2.9 & 2.7 &4.6 &  & 2.5 & 2.5 & 2.5 &4.2\\
	& 200 & 2.6 & 2.7 & 2.7 &4.3 &  & 2.5 & 2.5 & 2.5 &4.2\\
	\hline  
	\multirow{4}{*}{0.3} 
	& 50  & 2.9 & 0.9 & 2.0 &3.4 &  & 2.5 & 0.8 & 1.7 &2.9\\
	& 100 & 2.7 & 0.5 & 1.7 &3.0 &  & 2.5 & 0.5 & 1.6 &2.7\\
	& 150 & 2.6 & 0.4 & 1.7 &2.8 &  & 2.5 & 0.3 & 1.5 &2.6\\
	& 200 & 2.6 & 0.2 & 1.6 &2.7 &  & 2.5 & 0.2 & 1.5 &2.6\\
	\hline   
	\multirow{4}{*}{0.4} 
	& 50  & 2.9 & 0.3 & 1.7 &3.0 &  & 2.5 & 0.2 & 1.5 &2.6\\
	& 100 & 2.7 & 0.1 & 1.6 &2.7 &  & 2.5 & 0.1 & 1.5 &2.5\\
	& 150 & 2.6 & 0.0 & 1.6 &2.6 &  & 2.5 & 0.1 & 1.5 &2.5\\
	& 200 & 2.6 & 0.0 & 1.6 &2.6 &  & 2.5 & 0.1 & 1.5 &2.5\\
	\hline                     
	\multirow{4}{*}{0.6} 
	& 50  & 2.9 & 0.0 & 1.7 &2.9 &  & 2.5 & 0.1 & 1.5 &2.5\\
	& 100 & 2.7 & 0.0 & 1.6 &2.7 &  & 2.5 & 0.1 & 1.5 &2.5\\
	& 150 & 2.6 & 0.0 & 1.6 &2.6 &  & 2.5 & 0.1 & 1.5 &2.5\\
	& 200 & 2.6 & 0.0 & 1.6 &2.6 &  & 2.5 & 0.1 & 1.5 &2.5\\
\end{tabular}}
\end{table}

\renewcommand{\baselinestretch}{1.1}
\hspace{-2cm}
\begin{table}[h!tb] \centering
\caption{Empirical power (in \%, based on 3,000 repetitions) for $T_1,  T_{2, \Delta_0}(w)$ and the combined test $T_{c, \Delta_0} (w)$ with $w=1/4 $ or $  w_{\rm opt}$, where $\theta_0 = 0$, $\Delta^\star = 0.2$, $n_{1}: n_{0}: n_{e} = 2:1:3$,  $\sigma_1 = \sigma_0 = \sigma_{e} = 1$, and $\alpha=2.5\%$. In the table, we omit the $\Delta_0$ subscript for notational simplicity.}
\label{tab:simu empirical power}\vspace{3mm}
\tabcolsep= 3pt 
\resizebox{\textwidth}{!}{\begin{tabular}{cccccccccccccccccccccccccccc}
	& &
	\multicolumn{5}{c}{$\theta^\star = 0.2$}   & &
	\multicolumn{5}{c}{$\theta^\star = 0.3$} & &
	\multicolumn{5}{c}{$\theta^\star = 0.4$}\\
	\cmidrule{3-7} \cmidrule{9-13} \cmidrule{15-19}
	$\Delta_0$ &$n_1$ &
	$T_1$&
	$T_2(1/4)$ &  
	$T_2(w_{\rm opt})$ &     
	$T_c(1/4)$&   
	$T_c(w_{\rm opt})$ & &
	$T_1$&
	$T_2(1/4)$ & 
	$T_2(w_{\rm opt})$ &     
	$T_c(1/4)$&   
	$T_c(w_{\rm opt})$ & &
	$T_1$&
	$T_2(1/4)$ & 
	$T_2(w_{\rm opt})$ &     
	$T_c(1/4)$&   
	$T_c(w_{\rm opt})$\\     
	\hline
	\multirow{4}{*}{0.2} & 50  & 13.5 & 21.7 & 22.3 & 20.6 & 20.6 &  & 24.2 & 42.4 & 42.4 & 38.2 & 38.2 &  & 38.2 & 64.3 & 64.7 & 58.9 & 58.9 \\
	& 100 & 21.6 & 36.0   & 36.0   & 32.8 & 32.8 &  & 39.6 & 68.9 & 68.9 & 64.2 & 64.2 &  & 62.6 & 89.8 & 89.7 & 87.2 & 87.1 \\
	& 150 & 29.3 & 50.6 & 50.9 & 46.5 & 46.5 &  & 57.3 & 84.7 & 84.7 & 80.4 & 80.4 &  & 80.7 & 97.9 & 97.9 & 97.0   & 97.0   \\
	& 200 & 37.7 & 64.2 & 64.2 & 58.9 & 58.7 &  & 69.2 & 93.8 & 93.8 & 91.8 & 91.8 &  & 90.9 & 99.6 & 99.6 & 99.3 & 99.3 \\
	\hline
	\multirow{4}{*}{0.3} & 50  & 13.5 & 12.3 & 14.5 & 13.8 & 14.1 &  & 24.2 & 26.6 & 30.3 & 28.0   & 28.6 &  & 38.2 & 48.2 & 50.9 & 47.0   & 47.2 \\
	& 100 & 21.6 & 17.1 & 22.1 & 21.1 & 22.2 &  & 39.6 & 45.2 & 48.8 & 46.0   & 46.1 &  & 62.6 & 75.1 & 76.7 & 73.7 & 74.2 \\
	& 150 & 29.3 & 24.0   & 30.3 & 28.6 & 30.4 &  & 57.3 & 60.5 & 65.9 & 62.7 & 63.7 &  & 80.7 & 90.6 & 92.1 & 89.6 & 90.0   \\
	& 200 & 37.7 & 30.5 & 39.9 & 37.4 & 38.9 &  & 69.2 & 75.0   & 78.3 & 76.2 & 76.7 &  & 90.9 & 96.7 & 97.5 & 96.8 & 97.1 \\
	\hline
	\multirow{4}{*}{0.4} & 50  & 13.5 & 5.6  & 13.5 & 10.4 & 13.5 &  & 24.2 & 15.7 & 24.3 & 22.5 & 24.4 &  & 38.2 & 30.8 & 40.7 & 38.3 & 40.3 \\
	& 100 & 21.6 & 6.3  & 21.6 & 16.8 & 21.6 &  & 39.6 & 22.7 & 39.7 & 36.7 & 39.8 &  & 62.6 & 53.6 & 65.1 & 62.7 & 64.6 \\
	& 150 & 29.3 & 7.5  & 29.3 & 23.5 & 29.3 &  & 57.3 & 32.4 & 57.3 & 51.1 & 57.5 &  & 80.7 & 70.0   & 83.1 & 79.9 & 82.5 \\
	& 200 & 37.7 & 9.1  & 37.7 & 30.8 & 37.7 &  & 69.2 & 41.4 & 69.2 & 64.6 & 69.3 &  & 90.9 & 83.7 & 92.9 & 91.5 & 92.4 \\
	\hline
	\multirow{4}{*}{0.6} & 50  & 13.5 & 0.9  & 13.5 & 9.2  & 13.5 &  & 24.2 & 3.0    & 23.7 & 18.7 & 24.2 &  & 38.2 & 9.5  & 38.2 & 31.7 & 38.2 \\
	& 100 & 21.6 & 0.1  & 21.6 & 15.8 & 21.6 &  & 39.6 & 2.5  & 39.1 & 32.6 & 39.7 &  & 62.6 & 12.3 & 62.6 & 54.2 & 62.6 \\
	& 150 & 29.3 & 0.1  & 29.3 & 22.4 & 29.3 &  & 57.3 & 2.5  & 56.4 & 47.5 & 57.3 &  & 80.7 & 17.2 & 80.7 & 74.3 & 80.7 \\
	& 200 & 37.7 & 0.1  & 37.7 & 30.2 & 37.7 &  & 69.2 & 2.6  & 68.5 & 61.3 & 69.2 &  & 90.9 & 21.4 & 90.9 & 86.5 & 90.9\\
\end{tabular}}
\end{table}

\section{Proof of (\ref{eq: w opt})}
To maximize asymptotic power of $ T_{2, \Delta_0}(w)$ in  (\ref{eq: power2}), we find the $w$ that minimizes
$$
g(w)= \frac{(\theta_0 - \theta^\star )+ (1-w)(\Delta_0-\Delta^\star) }{\sqrt{ \pi_1^{-1}\sigma_1^2 +w^2  \pi_0^{-1}\sigma_0^2 +(1-w)^2n_r n_e^{-1} \sigma_e^2}}.
$$
To simplify notations, let 
$a = (\theta_0 - \theta^\star ), b =(\Delta_0-\Delta^\star), c = \pi_1^{-1}\sigma_1^2, d = \pi_0^{-1}\sigma_0^2, e = n_r n_e^{-1} \sigma_e^2$. When $w=1$, the $g(w)<0$ as $\theta^\star -\theta_0>0$, thus the $w$ should satisfy both $a+b(1-w)<0$ and maximize 
$$f(w) := g(w)^2 = \frac{(a+b(1-w))^2}{c+dw^2+e(1-w)^2}.$$
The derivative of $f(w)$ is 
\begin{align*}
f'(w) &= \frac{2(a+b(1-w))(-b)(c+dw^2+e(1-w)^2-(2dw+2e(w-1))(a+b(1-w))^2}{\left(c+dw^2+e(1-w)^2\right)^2}\\
&= \frac{-2(a+b(1-w))\left((bc-ae)+(ad+ae+bd)w\right)}{\left(c+dw^2+e(1-w)^2\right)^2}.
\end{align*}
The maxima $w\in [0,1]$ could only be among $w=\frac{ae-bc}{ad+ae+bd}$,  $w=0$, or $w=1$.\\
Since $a<0, b\geq 0, c, d, e>0$:\\
(1) when $ad+ae+bd>0$, that is, $-b/a = \frac{\Delta_0-\Delta^\star}{\theta^\star-\theta_0}> (d+e)/d$, we have $w=\frac{ae-bc}{ad+ae+bd}<0$, so it can't be a maxima. Also recall that we need to maintain $a+b(1-w)<0$ at the maxima. However when $w=0$, $a+b(1-w)=a+b>0$. Thus the maxima is $w=1$ in this case.\\
(2) when $ad+ae+bd<0$, that is, $-b/a = \frac{\Delta_0-\Delta^\star}{\theta^\star-\theta_0}< (d+e)/d.$
In this case, $f'(1)=\frac{-2a(bc+ad+bd)}{(c+d)^2}$ and $f'(0)=\frac{-2(a+b)(bc-ae)}{(c+e)^2}$. Further consider:\\
(2.1) If $-b/a < \frac{d}{c+d}$, i.e, $bc+ad+bd<0$, then $f'(1)=\frac{-2a(bc+ad+bd)}{(c+d)^2}<0$ concludes $w=1$ can't be a maxima. Also in this case $a+b<0$, so $f'(0)=\frac{-2(a+b)(bc-ae)}{(c+e)^2}>0$ concludes $w=0$ can't be a maxima. In this case the maxima is $w=\frac{ae-bc}{ad+ae+bd}$.\\
(2.2) If $-b/a >\frac{d}{c+d}$, then for $w=\frac{ae-bc}{ad+ae+bd}$ is not in $[0,1]$ as $1-w=\frac{ad+bd+bc}{ad+ae+bd}<0$. In this case the maxima would only be $w=0$ or $w=1$. Further assume: \\
(2.2.1) If $-b/a >1$, then $g(0)>0, g(1)<0$, then $w=1$ is the maxima.\\
(2.2.2) if $\frac{d}{c+d}< -b/a <1$, $g(0)<0, g(1)<0$. $f(0)=\frac{(a+b)^2}{c+e}$, $f(1) = \frac{a^2}{c+d}$. Thus we have $f(0)/f(1) = \left(\frac{a+b}{a}\right)^2 \frac{c+d}{c+e}< \left(\frac{a+b}{a}\right)^2 \frac{c+d}{c}<\frac{c}{c+d}<1$. The second last inequation holds as $(\frac{-c}{c+d})< -(a+b)/a <0$. In this case, the maxima is $w=1$.

In summary,\\
1. When $-b/a>\frac{d}{c+d}$, the maximum is achieved at $w=1$.\\
2. When $-b/a<\frac{d}{c+d}$, the maximum is achieved at $w=\frac{ae-bc}{ad+ae+bd}$. 

\end{document}